\documentclass[twocolumn,showpacs,preprintnumbers,amsmath,amssymb,prl]{revtex4}

\usepackage{graphicx}
\usepackage{dcolumn}
\usepackage{bm}
\usepackage{epsf,epsfig,latexsym}

\begin{document}
\title{Properties of the Initial Participant Matter Interaction Zone in 
Near Fermi-Energy Heavy Ion Collisions}
\author{J. Wang}
\thanks{Now at Institute of Modern Physics Chinese Academy of Science, Lanzhou 73, China}
\author{T. Keutgen}
\thanks{Now at FNRS and IPN, Universit\'e Catholique 
de Louvain, B-1348 Louvain-Neuve, Belgium}
\author{R. Wada}
\author{K. Hagel}
\author{S. Kowalski}
\thanks{Now at Institute of Physics, Silesia University, Katowice, Poland}
\author{T. Materna}
\author{L. Qin}
\author{Z. Chen}
\author{J. B. Natowitz}
\author{Y. G. Ma}
\thanks{on leave from Shanghai Institute of Nuclear Research,
Chinese Academy of Sciences, Shanghai 201800, China}
\author{M. Murray}
\thanks{Now at University of Kansas, Lawrence, Kansas 66045-7582}
\author{A. Keksis}
\author{E. Martin}
\author{A. Ruangma}
\author{D. V. Shetty}
\author{G. Souliotis}
\author{M. Veselsky}
\author{E. M. Winchester}
\author{S. J. Yennello}
\affiliation{Cyclotron Institute, Texas A\&M University, College Station, 
Texas 77843, USA}
\author{D. Fabris}
\author{M. Lunardon}
\author{S. Moretto}
\author{G. Nebbia}
\author{S. Pesente}
\author{V. Rizzi}
\author{G. Viesti}
\affiliation{INFN and Dipartimento di Fisica dell' Universit\'a di Padova, 
I-35131 Padova, Italy}
\author{M. Cinausero}
\author{G. Prete}
\affiliation{INFN, Laboratori Nazionali di Legnaro, I-35020 Legnaro, Italy}
\author{J. Cibor}
\affiliation{Institute of Nuclear Physics, ul. Radzikowskiego 152, 
PL-31-342 Krakow, Poland}
\author{Z. Majka}
\affiliation{Jagellonian University, M Smoluchowski Institute of Physics, 
PL-30059, Krakow, Poland}
\author{W. Zipper}
\affiliation{Institute of Physics, University of Silesia, PL-40007, 
Katowice, Poland}
\author{P. Staszel}
\affiliation{Jagellonian University, M Smoluchowski Institute of Physics, 
PL-30059, Krakow, Poland}
\author{Y. El Masri}
\affiliation{FNRS and IPN, Universit\'e Catholique de Louvain, 
B-1348 Louvain-Neuve, Belgium}
\author{R. Alfarro}
\author{A. Martinez-Davalos}
\author{A. Menchaca-Rocha}
\affiliation{Instituto de Fisica, Universidad National Autonoma de Mexico, 
Apactado Postal 20-364 01000, Mexico City, Mexico}
\collaboration {The NIMROD collaboration}
\noaffiliation
\author{~and~A.~Ono}
\affiliation{Department of Physics, Tohoku University, Sendai 980-8578, Japan}

\date{\today}

\begin{abstract}
The sizes, temperatures and free neutron to proton ratios of the initial 
interaction zones produced in the collisions of 40 MeV/nucleon 
$^{40}$Ar + $^{112}$Sn and 55 MeV/nucleon$^{27}$Al + $^{124}$Sn are derived 
using total detected neutron plus charged particle multiplicity as a 
measure of the impact parameter range and number of participant nucleons. 
The size of the initial interaction zone, determined from a coalescence 
model analysis, increases significantly with decreasing impact parameter. 
The temperatures and free neutron to proton ratios in the interaction zones 
are relatively similar for different impact parameter ranges and evolve in 
a similar fashion.
\end{abstract}

\maketitle

\section{Introduction}
For a collision between two heavy nuclei, measurements of emission 
cross-sections for early emitted nucleons and light clusters offer a means 
to probe the properties and evolution of the interaction region at early 
stages of the collision. Since light cluster production in such collisions 
reflects the particle-particle correlations within this interaction region, 
detection of a cluster can be viewed as a correlation measurement of its 
constituent particles in a bound state. Together with suitable application 
of a coalescence ansatz~\cite{csernai86,mekjian78,sato81,awes81_1,llope95}, 
this approach provides information which is complementary to that obtained in 
particle-particle correlation measurements that are well established in 
the nuclear context and have been applied in a wide range of 
studies~\cite{bauer92,ardouin97}. We have previously applied these techniques 
to obtain information on the early reaction dynamics and on the thermal 
evolution of the hot nuclei produced in near Fermi energy heavy ion 
collisions~\cite{cibor00,cibor01,hagel00,wang05_1,wang05_2}.  In this 
paper we report on the use of coalescence model analyses of light particle 
emission to probe the impact parameter dependence of the properties 
of the initial interaction zone and the evolution of participant matter 
produced in collisions   of 40 MeV/nucleon $^{40}$Ar + $^{112}$Sn and 
55 MeV/nucleon $^{27}$Al + $^{124}$Sn. We find that the size of the 
initial interaction zone increases significantly with decreasing 
impact parameter. The temperatures and free neutron to proton ratios 
in the interaction zones are relatively similar for different impact 
parameter ranges and they evolve in a similar fashion. 

\section{Experiment}
The reactions 40 MeV/nucleon $^{40}$Ar + $^{112}$Sn and 55 MeV/nucleon 
$^{27}$Al + $^{124}$Sn were studied at the K-500 super-conducting 
cyclotron facility at Texas A\&M University. For these 
studies we used the NIMROD detector array consisting of a $4\pi$ charged 
particle array inside a $4\pi$ neutron calorimeter~\cite{wada04,schmitt95}. 
The charged particle detector array of NIMROD includes 166 individual CsI 
detectors arranged in 12 rings in polar angles from $\sim4^o$ 
to $\sim160^o$. In these experiments Si-CsI telescopes 
were used to identify intermediate mass fragments (IMF). For the 
present experiment each forward ring also included two 
``super-telescopes'', each containing two Si $\Delta$E detectors 
and a CsI E detector and seven telescopes containing a single Si 
$\Delta$E detector and a CsI E detector.  Neutron multiplicity was 
measured with the $4\pi$ neutron detector surrounding 
the charged particle array. This detector is a neutron calorimeter 
filled with a Gadolinium-doped pseudocumene liquid scintillator. 
Thermalization and capture of emitted neutrons leads to scintillation 
which is observed with phototubes providing event by event determinations 
of neutron multiplicity but little information on neutron energies and 
angular distributions.  Further details on the detection system, 
energy calibrations and neutron calorimeter efficiency may be found 
in references~\onlinecite{wada04}.  During the experiment, data 
were taken employing two different trigger modes, one a minimum bias 
trigger in which at least one of the CsI detectors detected a particle 
and the other a high multiplicity trigger which 
required detected particles in 3 to 5 CsI 
detectors (depending upon the reaction studied). We have previously 
reported on excitation energy deposition and composite nucleus 
de-excitation in the most violent collisions observed for these 
systems~\cite{wang05_2}

\section{Data Analysis}
Many of the techniques applied in this analysis have been discussed 
previously in greater detail in 
references~\onlinecite{wang05_1,wang05_2,cibor00,hagel00,cibor01,wada04,ma05}. 
Only a brief summary of these techniques is included in the present work. 

For the reaction systems studied, an inspection of the two dimensional 
arrays depicting the detected correlation between charged particle 
multiplicity and neutron multiplicity in NIMROD reveals a distinct 
correlation in which increasing charged particle multiplicity is 
associated with increasing neutron multiplicity. 

Although there are significant multiplicity fluctuations reflecting both 
the competition between the different decay modes and the instrumental 
detection efficiencies, the data show that the total number of emitted 
particles can serve as a means for categorizing collisions according to 
overall degree of collision violence. Simulations with the AMD-V 
transport code~\cite{ono99}lead to similar conclusions. For the two 
reactions considered, the experimental distributions of total neutron 
plus charged particle multiplicity are shown in Figure 1. For the 
analysis reported in this paper we have used the total combined 
charged particle plus neutron multiplicities to select event classes 
for further analysis. For each of the reactions, events corresponding 
to four different regions of observed total detected neutron plus 
charged particle multiplicity were selected for analysis. These 
regions in total multiplicity are indicated by the dashed lines in Figure 1. 
\begin{figure}
\epsfig{file=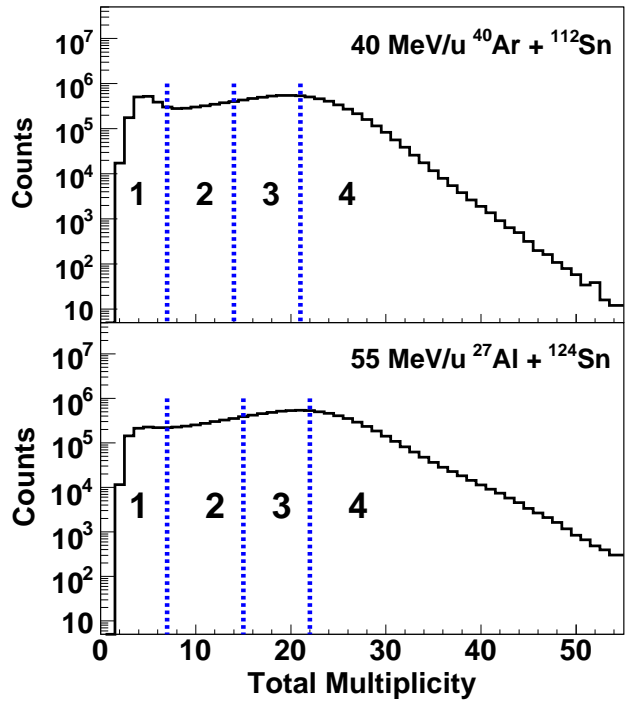,width=9.2cm,angle=0}
\caption{Total detected multiplicity of charged particles and  neutrons 
observed with the NIMROD detector.  Dashed lines indicate the multiplicity 
bins selected for  the analysis. (a) 40MeV/nucleon$^{40}$Ar + $^{112}$Sn  
(b) 55 MeV/nucleon$^{27}$Al + $^{124}$Sn}
\label{fig1}
\end{figure}

For the events in each selected multiplicity region we then carried out 
analyses using three-source fits to the observed energy and angular 
distributions of the light charged particles. The assumed sources 
were the PLF ( projectile-like fragment )  source, the target-like 
fragment source (TLF) and an intermediate velocity (IV) 
source~\cite{awes81_1,awes81_2,wang05_2,prindle98,wada89}.  From these 
fits we obtained parameters describing the ejectile spectra and 
multiplicities which can be associated to the three different sources. 
For the reactions studied, the spectral parameters for of 
$^{1}$H, $^{2}$H, $^{3}$H, $^{3}$He and $^{4}$He emission from the 
different  sources,  derived from the fits, follow the trends of 
earlier reported values at such projectile 
energies~\cite{awes81_1,awes81_2,wang05_2,prindle98,wada89,fox88}. 
The IV source slope parameters for $^{1}$H, $^{2}$H, $^{3}$H, 
$^{3}$He and $^{4}$He are characteristic of those for pre-equilibrium 
emission in this projectile energy 
range~\cite{awes81_2,wada89,wang05_2,prindle98,fox88,chiang80,cervesato92}.
Given the continuous dynamic evolution of the system, such source fits 
should be considered as providing only a schematic picture of the emission 
process. We have employed them to estimate the multiplicities and energies 
of ejectiles emitted at each stage of the reaction for each region of 
multiplicities. Both the mass of light ejectiles associated with the IV 
source and that associated with the TLF source increase monotonically with 
collision violence.

In the following we shall be particularly interested in the properties of 
the ejectiles from the IV source. To explore this part of the emission 
further, we have employed the Glauber model of reference~\onlinecite{glauber}
to estimate the number of participating nucleons corresponding 
to the different selected bins in total multiplicity. For this purpose a 
sharp cut-off approximation was employed to divide the results of the 
Glauber model calculation into four impact parameter bins, ranging from 
peripheral to central. These bins were matched to the bins employed for 
the experimental data by assuming that increasing violence corresponds 
to decreasing impact parameter and choosing the impact parameter ranges 
for each bin to assure that it contained an identical fraction of the 
total reaction cross-section to that of the corresponding experimental 
sample. For each bin the average number of participant nucleons, 
A$_{part}$, was then determined from the Glauber model calculation.  
In Figure 2 we present the relationship between the average numbers 
of participating nucleons and the yields of ejectiles for emission from 
the IV source. For both systems, the mass yield is seen to increase 
monotonically with A$_{part}$, confirming the strong correlation between 
the collision violence and the number of early emitted particles. This 
reflects the early collision dynamics within the initial interaction 
zone containing the participant matter from the two collision partners.

\begin{figure}
\epsfig{file=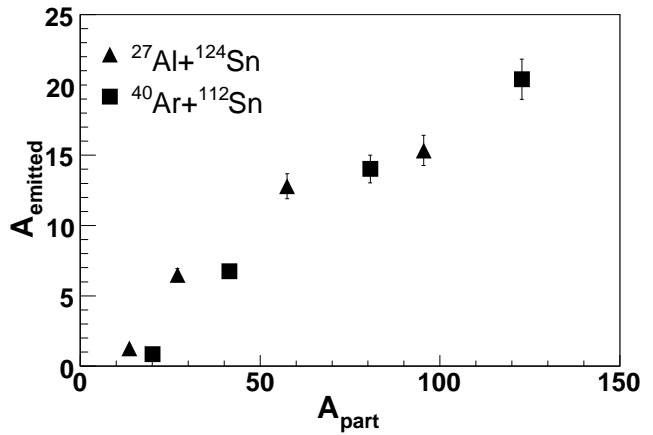,width=9.2cm,angle=0}
\caption{Mass emitted from the IV source as nucleons and light clusters 
plotted as a function of A$_{part}$ determined from Glauber model 
calculations. Results are presented for 40 MeV/nucleon $^{40}$Ar + $^{112}$Sn 
(solid squares) and 55 MeV/nucleon $^{27}$Al + $^{124}$Sn (solid diamonds).}
\label{fig2}
\end{figure}

In the following we attempt to probe further into the nature of this 
initial participant zone.

\section{Participant zone properties}

To probe the properties of the initial interaction zone, i.e., the sizes, 
temperatures and N/Z ratios of these zones, we have applied coalescence 
model analyses~\cite{csernai86,mekjian78,sato81}.  
In coalescence models the yields of ejected light clusters are directly 
related to the free nucleon yields. The phase space correlations which 
lead to cluster formation may be parameterized in terms of the momentum 
space volume within which the correlations between nucleons exist. This 
momentum space volume is assumed to be spherical with a radius of P$_{0}$. 
Analysis of the nucleon and cluster yields and extraction of P$_{0}$ 
provides information on the properties of the emission zone. 

To determine the coalescence parameter, P$_{0}$, in our energy range we have 
followed the Coulomb corrected coalescence model formalism of 
Awes~{\it et al.}~\cite{awes81_1} for which the laboratory frame differential 
yield for a cluster of Z protons and N neutrons, having  mass number A and 
a Coulomb-corrected energy per nucleon E$_{A}$ is:
\begin{eqnarray}
\frac{d^2N(Z,N,E_A)}{dE_Ad\Omega} =& 
\Big(\frac{N_t+N_p}{Z_t+Z_p}\Big)^N\frac{A^{-1}}{N!Z!}&  \nonumber \\
&\times\Big(\frac{\frac{4}{3}\pi P_0^3}{[2m^3(E-E_c)]
^{\frac{1}{2}}}\Big)^{A-1}&\nonumber\\
&\times\Big(\frac{d^2N(1,0,E)}{dEd\Omega}\Big)^A&
\label{eqnCoal}
\end{eqnarray}
where t denotes the target nucleus and p the projectile nucleus.

This cluster yield is directly related to the proton double differential 
yield at the same energy per nucleon, E, i.e., at the same velocity. The 
energy prior to Coulomb acceleration is obtained in the analysis by 
subtraction of the Coulomb barrier energy, E$_c$, derived from the source 
fits.  Since the system size may evolve during the particle emission stage we 
derive the parameter P$_{0}$ as a function of velocity as in previous 
works~\cite{cibor00,cibor01,hagel00,wang05_1,wang05_2}. The velocity we 
employ is the ``surface velocity'', V$_{surf}$, of the emitted particle, 
defined as the velocity of an emitted species at the nuclear surface, 
prior to acceleration in the Coulomb field~\cite{awes81_1}. To focus on 
the earlier evolution of the system we also subtracted the contributions 
from the TLF source from the total spectra. This was done using the 
experimentally determined fit parameters for the TLF source. Since the 
early emitted light particle energies are strongly correlated with 
emission times, and the evaporative or secondary emission contributions 
to the spectra are primarily at the lower kinetic energies, the yields 
of higher energy particles are relatively uncontaminated by later 
emission processes. To further focus on early particle emission we chose 
to work in the IV source frame and define V$_{surf}$ as the surface 
velocity in that frame. In that IV frame we selected nucleons and 
clusters emitted at mid-rapidity, i.e., at angles of 70 to 80 degrees 
in the IV source frame. In this way we attempted to isolate the emission 
associated with the IV source that occurs during the thermalization 
stage of the reaction by minimizing contributions from the PLF and 
TLF sources.

\section{$^3$H/$^3$He Ratios and $n/p$ Ratios}

\begin{figure}
\epsfig{file=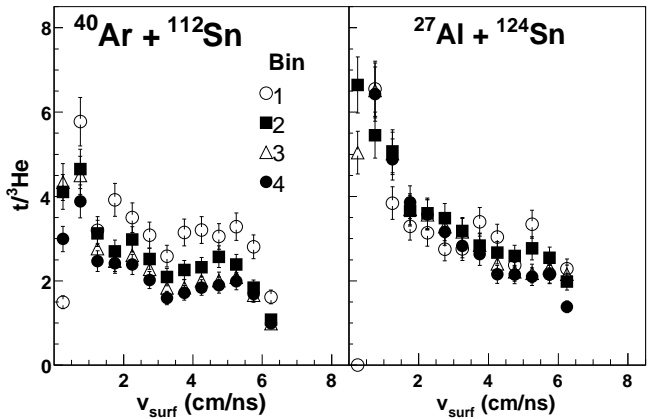,width=9.2cm,angle=0}
\caption{$^{3}$H/$^{3}$He ratios as a function of surface velocity. 
Symbols of open circles, solid squares, open triangles and solid 
circles  correspond respectively to progressively increasing total 
neutron plus charged particle multiplicity. (a) 40 MeV/nucleon 
$^{40}$Ar +$^{112}$Sn;  (b) 55 MeV/nucleon $^{27}$Al + $^{124}$Sn.
See text and Figure 1 for details.}
\label{fig3}
\end{figure}

As indicated by Equation~\ref{eqnCoal}, in this coalescence model the 
ratios of two isotopes which differ by one neutron are essentially 
determined by the ratio of  ``free nucleons ``in the coalescence volume. 
Thus, the free n/p ratio can be determined from a measurement of the 
$^3$H/$^3$He ratio~\cite{famiano06,keutgen}. In 
Figure~\ref{fig3} we present measured values of the $^3$H / $^3$He 
ratio as a function of V$_{surf}$.

Except at the very highest velocities, these ratios are seen to be 
significantly higher than the total N/Z ratios in the entrance channel 
(1.24 for $^{40}$Ar + $^{112}$Sn and 1.40 for $^{27}$Al + $^{124}$Sn). 
This is consistent with earlier results obtained by Albergo 
{\it et al.}~\cite{albergo85} who deduced significant 
free neutron excesses based on integrated yields observed in a variety 
of early intermediate energy experiments. Other recent work also 
results in large $^{3}$H/$^{3}$He ratios~\cite{hagel00,xu99,veselsky00}. 
It has been suggested that such observations provide evidence for a 
distillation leading to a nucleon vapor which is enriched in neutrons 
relative to a co-existing nuclear liquid in accordance with predictions 
of several theoretical studies~\cite{muller95,baran98}. 
However, Sobotka {\it et al.} have pointed out that symmetric cluster 
formation may play an important role in determining these 
ratios~\cite{sobotka97}.

\section{Coalescence Parameters, P$_0$}
\begin{figure}
\epsfig{file=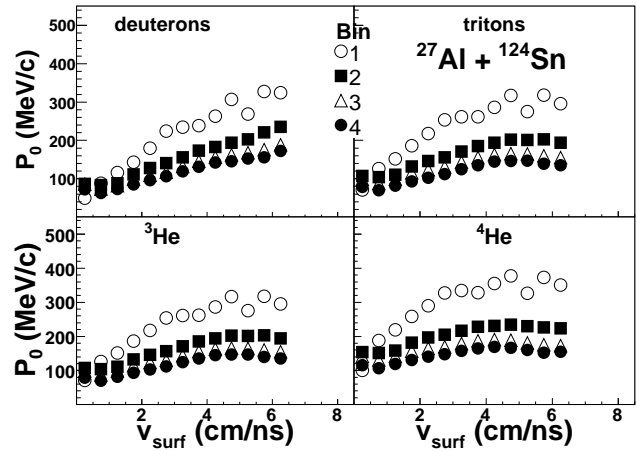,width=9.2cm,angle=0}
\caption{P$_0$ as a function of V$_{surf}$ for 55MeV/nucleon 
$^{ 27}$Ar + $^{124}$Sn. Results are presented for $^{2}$H (upper left), 
$^{3}$H (upper right), $^{3}$He (lower left) and $^{4}$He (lower right). 
In each case data are presented for four selected windows.  Symbols of 
open circles, solid squares, open triangles and solid circles  
correspond respectively to progressively increasing total neutron plus 
charged particle multiplicity See text.}
\label{fig4}
\end{figure}

\begin{figure}
\epsfig{file=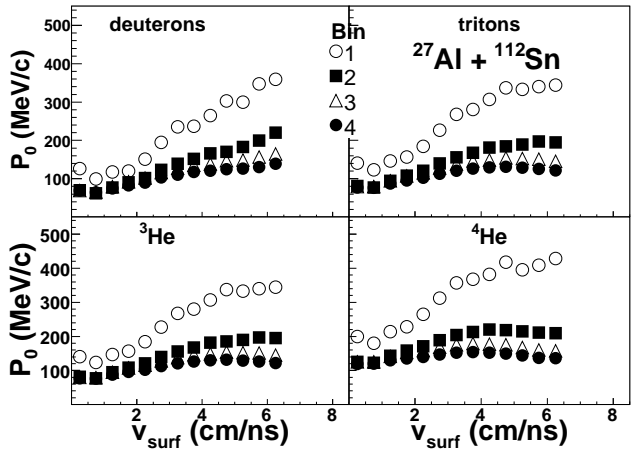,width=9.2cm,angle=0}
\caption{P$_0$ as a function of V$_{surf}$ for 40 MeV/nucleon 
$^{40}$Ar + $^{112}$Sn. Results are presented for $^{2}$H (upper left), 
$^{3}$H (upper right),$^{3}$He (lower left) and $^{4}$He (lower right). 
In each case data are presented for four selected windows. Symbols of 
open circles, solid squares, open triangles and solid circles 
correspond respectively to progressively increasing collision violence.
See text.}
\label{fig5}
\end{figure}

Using the observed $^3$H / $^3$He ratios to determine the n/p ratios 
required in equation 1, we have calculated the coalescence radius, P$_0$, 
as a function of V$_{surf}$. It should be noted that the method of 
derivation of the N/Z ratio from the $^3$H to $^3$He ratio leads 
to identical P$_0$ values for $^3$H and $^3$He. The results, 
presented in Figures~\ref{fig4} and~\ref{fig5} reveal that, for each light 
cluster, the derived values of P$_0$ decrease with decreasing V$_{surf}$ and 
also decrease with increasing total neutron plus charged particle multiplicity.
A closer inspection shows that the trend with surface velocity appears 
somewhat different for deuterons than for the other clusters. We also see a 
tendency for P$_0$ values for alpha particles in a given multiplicity bin 
to be larger than values for the other clusters. 

\section{Interaction Zone Sizes}
To extract nuclear size information from the P$_{0}$ measurements, the 
thermal coalescence model of Mekjian~\cite{mekjian78} has 
been employed. In the Mekjian model there is a direct relationship 
between the volume in momentum space and the coordinate space volume 
of the emitting system. In terms of the P$_{0}$ derived from 
Equation~\ref{eqnCoal} the relationship is: 
\begin{equation}
V=\Big(\Big(\frac{Z!N!A^3}{2^A}\Big)(2s+1)e^{\frac{E_0}{T}}\Big)^{\frac{1}{(A-1)}}
\frac{3h^3}{4\pi P_0^3}
\end{equation}

where Z, N and A have the same meaning  as in Equation 1, E$_{0}$ is the 
binding energy, s is the spin of the emitted cluster and T is the 
temperature. Thus in the coalescence model ansatz the volume of the 
emitting system can be derived from P$_{0}$. For this purpose, the 
temperature must be determined. Assuming a spherical shape of uniform 
density, the configuration space radius, R$_{0}$, may then be derived. 
This model assumes that both chemical and thermal equilibrium are achieved. 
Thus its applicability must be evaluated using a variety of experimental 
observables. In our previous work on similar systems, we have concluded 
that the data are consistent with achievement of such an equilibration, 
at least on a local basis~\cite{wang05_1,wang05_2}. This point is 
discussed further in the following section.

\section{Double Isotope Ratio Temperatures}
In an equilibrium model framework, the temperatures may be evaluated from 
double isotope yield ratio 
measurements~\cite{mekjian78,albergo85,wang05_1,wang05_2}. 
Using the same techniques as in references~\cite{wang05_1,wang05_2}
we have determined, the double isotope yield ratio temperatures, T$_{HHe}$, 
derived from the yields of $^2$H , $^3$H ,$^3$He and $^4$He 
clusters. This has been done as a function of ejectile velocity for each 
total multiplicity window for the two different systems under consideration. 
For particles emitted from a single source of temperature, T, and having 
a volume Maxwellian spectrum , $\epsilon^\frac{1}{2}e^{-\epsilon/T}$,where 
$\epsilon$ is the particle energy. The HHe double isotope yield ratio 
evaluated for particles of equal V$_{surf}$, is $(9/8)^{1/2}$ times the 
ratio derived from either the integrated particle yields or the yields 
at a given energy above the barrier~\cite{wang05_1}.  Thus  

\begin{equation}
T_{HHe} = \frac{14.3}{\ln{(\sqrt{(9/8)}(1.59 R_{Vsurf}))}}
\label{eqnT}
\end{equation}

If Y represents a cluster yield, 
R$_{Vsurf}$ = Y($^2$H)Y($^4$He)/Y($^3$H)Y($^3$He) for clusters with the 
same surface velocity and the constants 14.3 and 1.59 reflect binding 
energy, spin, masses and mass differences of the ejectiles. 
Equation~\ref{eqnT} differs from the usual formulation by a factor 
of $(9/8)^{1/2}$ appearing in the logarithm term in the 
denominator~\cite{albergo85}.  We present, in Figure 6, the resultant double 
isotope ratio temperatures, T$_{HHe}$, as a function of surface velocity. For 
the most violent collisions the temperature results have previously been 
reported~\cite{wang05_2}

\begin{figure}
\epsfig{file=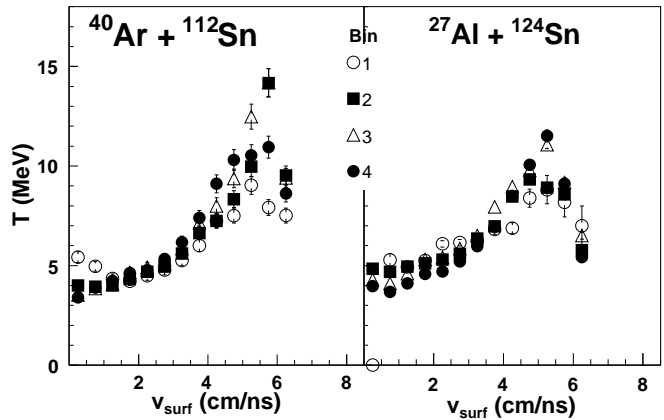,width=9.2cm,angle=0}
\caption{Evolution with surface velocity of the double isotope ratio 
temperature, T$_{HHe}$ for the two different reactions, 40~MeV/nucleon 
$^{40}$Ar + $^{112}$Sn -  and  55~MeV/nucleon $^{27}$Al + $^{124}$Sn. 
Symbols of open circles, solid squares, open triangles and solid 
circles  correspond respectively to progressively increasing total 
neutron plus charged particle multiplicity.  See text.}
\label{fig6}
\end{figure}

In Figure~\ref{fig6} we see that as V$_{surf}$ decreases from the highest 
V$_{surf}$ sampled, i.e., as reaction time increases, each of the 
temperature evolution curves exhibits a maximum and then decreases. 
Maximum temperatures of 8-14 MeV are observed.The trends in Figure~\ref{fig6}
are very similar to those reported for previous measurements of the 
temperature evolution in the reactions of 26-47 MeV/nucleon projectiles 
with various targets~\cite{cibor00,hagel00,wang05_1,wang05_2}.  In those works 
the correlation of decreasing surface velocity with increasing 
emission time is discussed. In reference~\onlinecite{wang05_1} the peaks 
in the temperature at surface velocities near 6 cm/ns were interpreted 
as corresponding to times in the range of  95 to 110 fm/c, depending 
upon reaction system.  After that time the temperature decreases   
monotonically with decreasing surface velocity. The AMD-V model 
calculations~\cite{ono99,wada04,wang05_1,wang05_2} for those systems indicate 
a significant slowing in the rate of change of the ejectile  kinetic 
energy near a velocity of 3.5 cm/ns, signaling  the end of the IV 
(or pre-equilibrium) emission stages and entry into the region of 
slower nuclear de-excitation modes, i.e., evaporation, fission and/or 
fragmentation. At that point the sensitivity of the emission energy to 
time is significantly reduced. Consequently, we take the temperature at 
the time corresponding to the velocity of 3.5 cm/ns to be that of the 
hot nucleus at the beginning of the final statistical emission stage 
(appropriate to initial emission from the TLF source.)  At that point 
the corresponding T$_{HHe}$ temperatures are near 6 MeV and thus very 
similar to the limiting temperatures previously derived from a 
systematic investigation of caloric curve measurements~\cite{natowitz02_1}, 
in this mass region. In reference~\cite{wang05_1,wang05_2} it is concluded 
that, for velocities below those corresponding to the peaks in the temperature 
curves, the temperature data are consistent with the achievement of 
chemical and thermal equilibration, at least as sampled on a local basis.

\section{Emission Zone Radii}
\begin{figure}
\epsfig{file=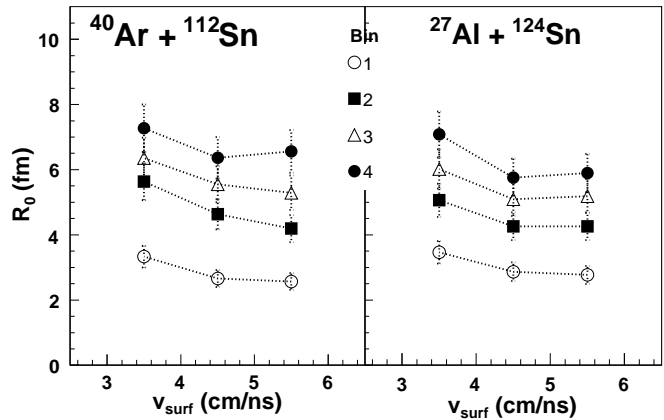,width=9.2cm,angle=0}
\caption{Interaction zone radii as a function of V$_{surf}$. Values averaged 
over bins of 1 cm/ns in  V$_{surf}$ are presented for four different windows 
of total neutron plus charged particle multiplicity for the systems 
(a) 40~MeV/nucleon $^{40}$Ar + $^{112}$Sn and (b) 55~MeV/nucleon 
$^{27}$Al + $^{124}$Sn. Symbols of open circles, solid squares, 
open triangles and solid circles correspond respectively to progressively 
increasing total neutron plus charged particle multiplicity.}
\label{fig7}
\end{figure}

In Figure~\ref{fig7}, average values of R$_0$ for the two systems, 
obtained from the four different ejectiles, are presented for the four 
different windows of total neutron plus charged particle multiplicity. 
Over the range of V$_{surf}$ from 3 to 6 cm/ns these values are averaged 
for 1 cm/ns intervals. Here the observed trends in P$_0$ result in a 
significant increase in R$_0$ with increasing total neutron plus 
charged particle multiplicity. The derived values of R$_0$ are seen 
to range from $\sim2.5$ to $\sim7.5$ fermis and be rather similar for the two 
systems. For the most violent collisions, values of R$_0$ at the 
highest surface velocities are 6 to 6.5 fermis, close to the expectations 
for the equivalent sharp cut-off radius of normal density nuclei with 
total mass numbers equal to those of the entrance channel, A=151 or 
152~\cite{preston62}. An increase of R$_0$ with decreasing velocity is 
seen for the the different impact parameter windows, The fractional 
changes for different windows are quite similar.

\begin{figure}
\epsfig{file=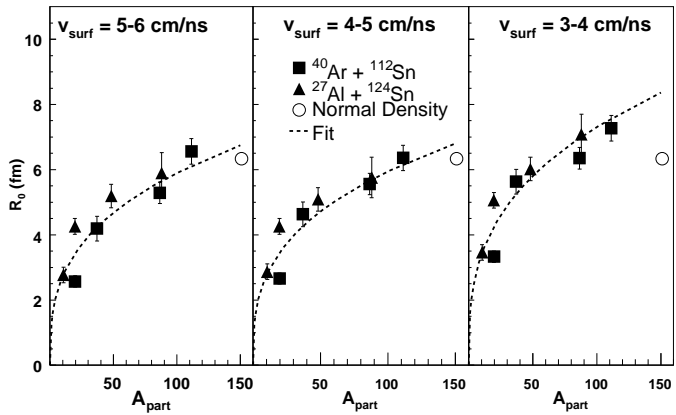,width=9.2cm,angle=0}
\caption{Emission zone radii vs A$^\prime_{part}$ (see text) for three 
different windows on V$_{surf}$. Results for both reactions are presented as 
average values of R0 for three bins of 1 cm/ns in V$_{surf}$ in the range of 
3 to 6~cm/ns.  Average values of V$_{surf}$ are (a) 5.5 cm/ns, 
(b) 4.5 cm/ns and (c,) 3.5 cm/ns. Symbols are solid triangles - 
55 MeV/nucleon$^{27}$Al + $^{124}$Sn and solid squares - 
40MeV/nucleon $^{40}$Ar + $^{112}$Sn.  For comparison, the equivalent 
sharp cut off radius for a normal density nucleus with A = 151 is 
indicated by the open circles in each part of the figure.}
\label{fig8}
\end{figure}

In order to further evaluate the evolution of the interaction zone size, 
we have explored the correspondence between interaction zone size and 
the number of participant nucleons. For this purpose we have adopted a 
somewhat different estimate of the later quantity by relaxing the 
sharp-cut-off approximation of the Glauber model estimate. As previous 
calculations employing the AMD model of Ono {\it et al.}~\cite{ono99}, 
had been found to reproduce well a variety of experimental observables 
for similar systems, we filtered results of AMD calculations for the 
two systems using the same conditions as employed for the experimental 
data. Not unexpectedly these results indicated that the impact 
parameter ranges selected by the adopted windows in total charged 
particle plus neutron multiplicity are not as sharply defined as 
those used to determine the average A$^\prime_{part}$ of the Glauber model. 
We then revised these estimates of A$^\prime_{part}$ by weighting the 
Glauber model results by the derived AMD impact parameter distributions. 
This procedure results in estimates of the participating nucleon 
numbers which are ~ 10\% lower than those obtained with the sharp 
cut-off assumption. In Figure~\ref{fig8} (a)-(c) the derived radii are plotted 
against these refined estimates, designated A$^\prime_{part}$. For comparison, 
fits of the function $R_0 = r_0(A_{part}^{\prime 1/3})$ are also shown. 
This function fits the data reasonably. The values of $r_0$ extracted 
from these fits increase with decreasing surface velocity. They are 
1.27fm, 1.28 fm and 1.51 fm respectively for the 5-6, 4-5 and 3-4 cm/ns 
windows. The scaling clearly implies that the size of the zone being 
sampled is proportional to A$^\prime_{part}$. If the zone contains A$^\prime_{part}$ 
nucleons and has a spherical shape, a comparison of the r$_0$values 
with the equivalent uniform radius parameter for normal density nuclei 
with A=151~\cite{preston62}, indicated by an open circle in 
each part of Figure~\ref{fig8}, would suggest corresponding average densities 
decreasing from $0.85\rho_0$ to $0.50\rho_0$ (where $\rho_0$ is the normal 
ground state nuclear density)as the system evolves and V$_{surf}$ 
decreases. Such average density estimates are close to those derived 
from a Fermi gas model analysis of caloric curves for similar 
systems~\cite{natowitz02_2}.
 
However, given that 1) many collisions should be Pauli blocked, that 2) 
some nucleons are emitted during this process and that 3) thermalization 
is occurring, it is difficult to extract precise information on the 
densities. Indeed, the AMD calculations for these and similar systems 
clearly indicate large density fluctuations with fragments of normal 
density imbedded in a lower density medium of nucleons and smaller 
clusters, a nuclear gas~\cite{ono99,wada04}. If the nuclear 
gas is equated to the early emitted particles from the  IV source, the 
observed scaling with A$_{part}^{\prime 1/3}$ may simply  
reflect the direct dependence of the total mass of the emitted particles 
on  the number of participant nucleons which is seen in Figure 2. In 
such a case, the interaction zone radii extracted from the coalescence 
model might better be viewed as those characterizing the nuclear gas 
and the densities derived from the ratios of the number of gas nucleons 
to the interaction zone radii would near 0.1$_{0}$. If this latter 
interpretation proves to be correct, it may be possible to employ 
analyses of such reaction data to test theoretical predictions of the 
properties of  low density nuclear gases~\cite{beyer00,shen98,horowitz05} 
in greater detail.  

\section{Summary and Conclusions}
The sizes, temperatures and free neutron to proton ratios of the initial 
interaction zones produced in the collisions of 40 MeV/nucleon 
$^{40}$Ar + $^{112}$Sn and 55 MeV/nucleon $^{27}$Al + $^{124}$Sn are 
derived for different degrees of collision violence, selected by gating 
on the total multiplicity of emitted neutrons and charged particles. 
The size of the initial interaction zone, derived from coalescence 
model analyses, increases significantly with total neutron plus 
charged particle multiplicity ($\sim$centrality). The temperatures and 
the free neutron to proton ratios in these zones exhibit very 
similar evolutions with decreasing surface velocity. The similar 
behavior of these observables with surface velocity for the different 
bins indicates that the thermal and chemical properties of the participant 
zones are very similar from the most peripheral to the most central 
collisions. The interaction zone radii obtained from the coalescence 
analyses have been found to correlate well with the number of nucleons 
in the participant matter region. The interpretation of this correlation 
is discussed and it is suggested that measurements of the type reported 
here may allow more detailed investigation of the properties of low 
density nuclear gases, a topic of both nuclear and astrophysical 
interest~\cite{beyer00,shen98,horowitz05}.  We are currently exploring this 
possibility.

\section{Acknowledgements}
This work was supported by the United States Department of Energy under 
Grant \# DE-FG03- 93ER40773 and by The Robert A. Welch Foundation under 
Grants \# A0330 and A-1266.

\end{document}